# Verifying Fixed-Point Digital Filters using SMT-Based Bounded Model Checking

Renato B. Abreu, Lucas Cordeiro, and Eddie B. L. Filho

*Abstract*—The implementation of digital filters in processors based on fixed-point arithmetic can lead to problems related to the finite word-length. In particular, the processing of signals in such filters can produce overflows and unwanted noise caused by quantization and round off effect during the accumulative addition and multiplication operations. In this paper, we describe a new approach to verify digital filters using an off-the-shelf bounded model checker called ESBMC, which supports full C/C++ and is based on satisfiability modulo theories solvers. In particular, we are able to verify the occurrence of overflows, limit cycles, and time constraints based on a discrete-time model implemented in C. The experiments show that the proposed approach can be used to verify potential problems in fixed-point implementation of digital filters and it can thus be effective in finding realistic design errors.

## I. INTRODUCTION

Digital filters have been widely used for digital signal processing applications due to its reduced computational complexity and flexibility enabled by available digital signal processors (DSPs) and field programmable gate arrays (FPGAs). Recently, the availability of floating-point processors has substantially grown, but the reduced cost and the high speed of fixed-point processors still make it the choice for the embedded digital filters projects. However, fixed-point implementation leads to quantization nonlinearities, round off errors, and overflows caused by consecutives multiplications and additions operations using finite word-length; and these may affect the desired behavior of the filter. As an example, for direct form realizations, only a small change on filter coefficients due to parameter quantization can result in a large change in the location of the poles and zeros of the system [1]. In recursive digital filters, most known as infinite impulse response (IIR) filters, can present serious oscillations in the output even for a zero input signal, which is a phenomenon known as *limit cycle*. Finite impulse response (FIR) filters do not suffer from such limit cycle effects, but they may have other issues caused by the finite word-length limitations (e.g., overflows). There are many studies about the quantization and limit cycle in digital filters, and ways to reduce its effects, as previously reported in [2], [3].

Apart from that, an important property to implement a digital filter for real-time applications is its time constraint. Modern microcontrollers and DSPs allow programming in high-level languages such as C. The filter program is compiled to low level instructions that consume clock cycles to be processed. That processing time must meet some constraints according to the system sample frequency and available buffer. Normally, the filter designers employ advanced tools to define filter parameters according to the desired operation in time and frequency domains, and use simulation software to validate their behavior under extensive tests. In most cases, they consider floating-point arithmetic in calculations. There are a few tools to simulate systems using fixed-point arithmetic [4], [5]. Search algorithms to determine the minimum bound of the word-length are also presented in [6] in which the authors adopt a simulation-based approach. However, testing and simulation can lead to a limited number of scenarios and inputs in the system, which thus do not exploit all possible behaviors that the system can exhibit. Hence, only frequency domain graphical analysis and simulation might not be sufficient to conclude about possible problems related to finite word-length implementation as well as time constraints of the filters.

Recently, some alternative technique has been proposed for the verification of fixed-point implementations of IIR digital filters, which is based on bounded model checking (BMC) and suggests the use of modern satisfiability modulo theory (SMT) solvers [7]. The main idea behind SMT-Based BMC is to consider counterexamples of a particular length $k$ and generate a first-order logic formula that is satisfiable if and only if such a counterexample exists [8]. In this paper, we describe the use of a general purpose SMT-based bounded model checker for embedded C/C++ software to verify potential issues caused by fixed-point arithmetic on recursive filters; it makes two major contributions. First, we consider the processing time of operations during the filter function unrolling to check for the maximum acceptable time of the filter operations. Second, we exploit BMC to verify the actual C code of the digital filter that is intended to be embedded into micro-controllers and DSPs; this is closer to the real implementation where specific C constructs (e.g., pointer arithmetic and comparisons) are used to implement the digital filter. Last but not least, the application of SMT-based BMC to digital filters might not be well known amongst DSP developers and so this work can potentially add value to them.

## II. FIXED-POINT FILTERS REALIZATION

Digital filters can be defined as linear time-invariant discrete-time systems described by a difference equation as:

$$y(n) = -\sum_{k=1}^{N} a_k\, y(n-k) + \sum_{k=0}^{M} b_k\, x(n-k) \quad (1)$$

where *y(n)* is the output in instant *n*, *y(n-k)* is the output $k$ steps in the past, *x(n-k)* are the inputs $k$ steps in the past, $a_k$ are the coefficients for the past outputs, $b_k$ are the coefficients for the inputs, *N* is the feedback filter order, and *M* is the feedforward filter order. The design of a digital filter mainly consists of finding the values of the coefficients $a_k$ and $b_k$ that produce the expected frequency response. The filters are usually classified

Renato B. Abreu[1,2], Lucas Cordeiro[2], and Eddie B. L. Filho[2,3] ¸ [1] Nokia Institute of Technology; [2] Department of Electronics and Computing, Federal University of Amazonas; and [3] Science, Technology, and Innovation Center for the Manaus Industrial Pole, Manaus-AM, Brazil, E-mails: renato.abreu@indt.org.br, lucascordeiro@ufam.edu.br, eddie@ctpim.org.br.

according to their ideal frequency domain characteristics. It is out of the scope of this paper to show IIR and FIR filters design methods. This is a huge topic that is covered in standard digital signal processing books [1], [9]. There are many ways to implement (1) in hardware, or in software on a programmable digital computer depending on the realization structure of the system. The commonly known Direct Form I realization for IIR systems is shown in Fig. 1. For the demonstration of the proposed method, Direct Form I structure was chosen to be implemented in the C language.

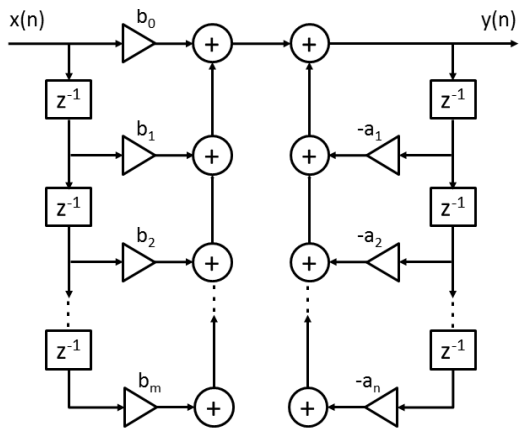

Figure 1. Direct Form I structure of IIR filter.

In the realization of fixed-point digital filters, the coefficients and the results of intermediate computations suffer the effect of quantization and round off errors. Here, we have considered the round off quantizer Q(x); for this quantizer, the maximum error caused by rounding is $2^{-b-1}$, where $b$ is the number of bits that belongs to the fractional part. In a case where the result from an addition or multiplication exceeds the amount of bits available for the number representation, we say that there is an *overflow*. For the limit cycle verification, we allow the overflows to naturally happen. We have thus considered the two's complement arithmetic so that when the overflow occurs, the result will wrap around. Fig. 2 shows the behavior of the round off quantizer and the effect of the two's complement overflow wrapping around.

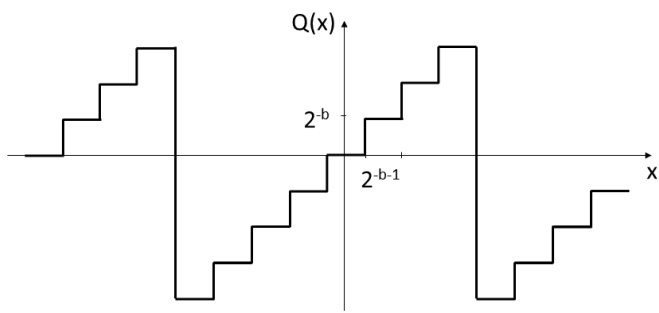

Figure 2. Round off quantizer of *b* bits with wrap around overflow.

To obtain a realistic model of the finite precision system, we consider the quantization of each numeric value in the system including inputs, coefficients, and results of arithmetic operations. Fig. 3 shows this model for a single-pole filter.

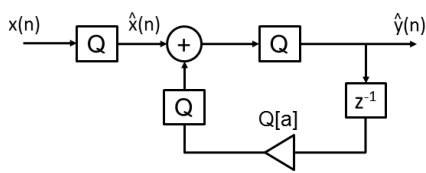

Figure 3. Realistic model of a single-pole quantized filter.

Here, we represent numbers in fixed-point format using a pair of digits separated by a decimal point. The digits to the left and right represent the integer and fractional parts, respectively. We use the two's complement to represent signed number in fixed-point processors. In this system, the real number $X$ described by the $\langle k, l \rangle$ fixed-point position number ($b_{k-1}\ b_{k-2}\ ...\ b_1\ b_0 \cdot b_{-1}\ b_{-2}\ ...\ b_{-l}$) can be represented as:

$$X = -b_{k-1}2^{k-1} + \sum_{i=k-2}^{-l} b_i 2^i \qquad (2)$$

The most significant bit $-b_{k-1}$ is used for the sign. Thus, the maximum value representable by a number that consists of an integer part with $k$ bits and a fractional part with $l$ bits is $2^{k-1} - 2^{-l}$, and the minimum value is $-2^{k-1}$. The quantizer represented in Fig. 3 by the block Q rounds the numbers inside this range. If a number does not fit in this interval, then it indicates an overflow. During the verification of the filter, we check the overflow as a failure in the system, or the quantizer wraps around the result (as shown in Fig. 2).

### III. SMT-BASED BMC OF DIGITAL FILTERS

The basic idea of BMC is to check (the negation of) a given property at a given depth: given a transition system $M$, a property $\phi$, and a bound $k$, BMC unrolls the system $k$ times and translates it into a verification condition (VC) $\psi$ such that $\psi$ is satisfiable if and only if $\phi$ has a counterexample of depth less than or equal to $k$. Standard SMT solvers can be used to check whether $\psi$ is satisfiable. In BMC of digital filters, the bound $k$ limits the number of loop iterations and recursive calls in the program. BMC thus generates VCs that reflect the exact path in which a statement is executed, the context in which a given function is called, and the bit-accurate representation of the expressions [8]. Proving the validity of the VCs arising from programs remains a major performance bottleneck, despite attempts to cope with increasing system complexity by applying SMT solvers. In this work, we used the Efficient SMT-Based Bounded Model Checker (ESBMC) tool as the verification engine since it was the most efficient BMC tool in the last two software verification competitions [10], [11]. In ESBMC, the associated SMT-based BMC problem is formulated by constructing the following logical formula:

$$\psi_K = I(s_0) \wedge \bigvee_{i=0}^{k} \bigwedge_{j=0}^{i-1} \gamma(s_j, s_{j+1}) \wedge \overline{\phi(s_i)} \qquad (3)$$

Here, $\phi$ is a safety property (e.g., overflow), $I$ is the set of initial states of $M$, and $\gamma(s_j, s_{j+1})$ is the transition relation of $M$ between time steps $j$ and $j+1$. Hence, $I(s_0) \wedge \bigwedge_{j=0}^{i-1}(s_j, s_{j+1})$ represents the executions of $M$ of length $i$. The above VC $\psi_k$ can be satisfied if and only if for some $i \leq k$ there exists a reachable state at time step $i$ in which $\phi$ is violated. If (3) is satisfiable, then the SMT solver provides a satisfying assignment, from which we can extract the values of the program variables to construct a counterexample. A counterexample for a property $\phi$ is then defined as a sequence of states $s_0, s_1, ..., s_k$ with $s_0 \in S_0, s_k \in S$, and $\gamma(s_j, s_{j+1})$ for $0 \leq j < k$. If (3) is unsatisfiable, then we can conclude that no error state is reachable in $k$ steps or less.

In this work, we propose the following steps for the design and verification of a digital filter. First, we design the filter parameters using the preferred methods (cf. [1], [9]) and tools (cf. [12]). After that, we estimate the output range for a given input range to define the word-length to represent the fixed-point numbers. Once we define the word-length, we input the respective design parameters into the C filter model and then perform time analysis of filter operations using specific microprocessor architecture. Finally, we add assertions to the given C model to check for properties related to time constraints, under and overflow, and limit cycle. If we find an under and overflow, or a limit cycle violation, then we increase the word-length and call the verification engine again. If we find a time constraint violation, then it indicates that we have to decrease the word-length to improve the performance.

*A. Arithmetic Underflow and Overflow Verification*

During the design of a fixed-point filter, one needs to specify the number of integer and fractional bits. Firstly, one needs to estimate the output range of the filter for a given input range; this estimation is typically based on analytical- or simulation-based approaches. Several authors have proposed techniques to find the word-length for the coefficients of digital filters as in [13], [14]. However, to detect overflows in a digital filter with a given fixed-point word-length and expected input range, we add assertions into the quantizer block and configure the verification engine to use non-deterministic inputs in the specified range. For any result of addition or multiplication in the filter operation, if there exists a value that exceeds the range representable by the fixed-point, an assert statement detects it as an underflow or overflow violation. We generate a literal $l_{resp\_op}$ to represent the validity of each addition and multiplication operation with the following constraint:

$$l_{resp\_op} \Leftrightarrow (MIN \leq FP) \land (FP \leq MAX) \quad (4)$$

where *FP* is the fixed-point approximation for the result of the adders and multipliers; *MIN* and *MAX* are the minimum and maximum values that are representable for the given fixed-point bit format (as we previously described in Section II). As a running example, let us consider a single pole system described by the following difference equation [1]:

$$y(n) = -a\, y(n-1) + x(n) \quad (5)$$

This is a *bounded-input bounded-output* (BIBO) stable system in which the output is limited in amplitude to:

$$|y(n)| \leq x_{max} \sum_{k=-\infty}^{\infty} |h_k| \quad (6)$$

where $x_{max}$ is the maximum input value and $h_k$ is the impulse response of the system. For the Eq. (5) with $a = -1/2$, it can be shown that the summation of the norm of the impulse response converges to 2 using geometric series. For this particular example, if we consider an input in the range [-1, 1], the output will be [-2, 2] (i.e., we simply multiply the input range by $\sum |h_k|$). If we have this for the implementation, one could choose to represent the fixed-point number using 2 bits for integer part including the sign and 4 bits for the fractional part. The resulting range for this particular format is [-2, 1.9375], with an error of ±0.03125.

Using the proposed method, we apply the coefficients of (5) to the filter implemented in the C language and then define the number of bits for the integer and the fractional parts of the fixed-point number. If we run the verification engine by taking into account the input range [-1, 1], then it shows a counterexample in which the system gets an overflow for a particular sequence of inputs. It can be easily shown that an input sequence x = {1, 1, 1, 1, 1, 1} leads to an overflow in the output as shown in Table I.

TABLE I. EXAMPLE OF OVERFLOW IN FIXED-POINT FILTER

| n | 1 | 2 | 3 | 4 | 5 | 6 |
|---|---|---|---|---|---|---|
| x(n) | 1 | 1 | 1 | 1 | 1 | 1 |
| y(n) | 1 | 1.5000 | 1.7500 | 1.8750 | 1.9375 | **1.96875**[*] |

[*]. Considering 4 bits to fractional part, value is out of range [-2, 1.9375]

For this particular case, one could easily infer about the overflow by analyzing the impulse response summation or by simulating a constant step input. However, for high order systems, it can be difficult to precisely evaluate about the impulse response infinite summation or find an input sequence that leads to overflow, as also observed by [7]; this thus motivates the application of BMC to digital filters.

*B. Limit cycle Verification*

In an ideal stable filter, the output should asymptotically approach a steady-state level determined by the filter transfer function [15]. The limit cycle can manifest either as a steady oscillation or as a nonzero level in the output, even for a zero level input. This effect is caused by the round off errors and overflows during the filter operation. To verify the presence of limit cycle in a particular fixed-point filter realization, we configure the quantizer block routine by setting a flag variable on it to enable the wrap around on overflows. The expected behavior will be as shown in Fig. 2, which means that the verification engine is not expected to detect the overflow failures as in the previous case. Additionally, we configure the filter to use a zero input signal and a non-deterministic initial state for the previous outputs. We thus unroll the filter execution for a bounded number of entries and add an assert statement to detect a failure if a set of previous outputs states (that repeats during the zero-input response) is found. Note that this method is slightly different from that presented by Cox et al. [7], which aims at finding a limit cycle by comparing a window of the output with another window of the output within a bounded number of steps later.

As an example, let us consider the same system described by the difference equation in (5). Here, we also model the system using 2 bits for the integer part and 4 bits for the fractional part as in the previous case, but now we set a zero input signal instead. If we execute the verification engine for the implemented model, then it finds a particular initial condition that leads the system to a limit cycle. In Table II, we present the response of the system for that particular condition. Note that the columns $y_2$ and $y_{10}$ represent the filter response in binary and decimal format, respectively. Due to the rounding on the fractional part of the fixed-point number, we can see in Table II that for $a = 0.5$ the output starts to repeat after $n = 2$. Similarly, for $a = -0.5$, we can see that it keeps a nonzero steady-state value instead of decaying to zero.

TABLE II. LIMIT CYCLES FOR SINGLE POLE FILTER

| $a = 0.5_{10} = 0.1000_2$ | | | $a = -0.5_{10} = 1.1000_2$ | | |
|---|---|---|---|---|---|
| n | $y_2$ | $y_{10}$ | n | $y_2$ | $y_{10}$ |
| -1 | 0.0010 | 0.125[*] | -1 | 0.0010 | 0.125[*] |
| 0 | 1.0001 | -0.0625 | 0 | 0.0001 | 0.0625 |
| 1 | 0.0001 | 0.0625 | 1 | 0.0001 | 0.0625 |
| 2 | 1.0001 | -0.0625 | 2 | 0.0001 | 0.0625 |
| 3 | 0.0001 | 0.0625 | 3 | 0.0001 | 0.0625 |

[*]. Initial condition found as counterexample

## C. Time Constraints Verification

There are efficient structures for the implementation of filters such as Lattice form and filtering methods based on the Fast Fourier transform [1]. These methods aim to reduce the number of arithmetic operations and computations costs. However, the time-domain convolution methods based on direct forms are still prevalent, both in hardware and in software implementations, due to its simplicity. In real-time applications, filter receives data at the same rate it processes and outputs it. As a result, the verification of time constraints becomes necessary, especially in high order filters, which present more arithmetic operations and higher group delays.

In the proposed approach, we use the filter model to check and assert about the maximum acceptable time for the filter operations to be processed. As an example, we implemented an IIR filter function and compiled it to run on a MSP430G2231, which is an ultra-low-power 16-bit RISC CPU based microcontroller [16]. If we configure the compiler to generate the assembly file and merge it together with C source code, then we can perform a worse case execution time (WCET) analysis in the IIR function. As an example, the code fragment shown in Fig. 4 is used to perform the multiplication of the $b_k$ coefficients with the previous entries in Eq. (1). Fig. 5 shows the code of Fig. 4 converted into some assembly instructions using the compiler CCS v4 [17]:

```
sum += *b_ptr++ * *x_ptr--;
```

Figure 4. C code fragment of the digital filter.

```
MOV.W @r9+,r12      5 cycles
MOV.W @r9+,r13      5 cycles
SUB.W #4,r10        5 cycles
MOV.W 4(r10),r14    3 cycles
MOV.W 6(r10),r15    3 cycles
CALL  #__fs_mpy     5 cycles
MOV.W r7,r14        1 cycle
MOV.W r8,r15        1 cycle
CALL  #__fs_add     5 cycles
MOV.W r12,r7        1 cycle
MOV.W r13,r8        1 cycle
```

Figure 5. Assembly instructions of the code fragment shown in Fig. 4.

We can see that each instruction can take a different number of clock cycles to execute. Based on that information, we can compute how many clock cycles will be needed for each operation. For the MSP430G2231, the internal frequency is up to 16 MHz that gives a cycle time of 62.5 ns. Once we have the total time of the instructions, then we can use this to increment a timer variable and add an assert statement to detect any violation of time constraints. The value of the constraint can be easily estimated based on the sample rate of the system. If the system operates using a sample rate of 48 KHz (which is commonly used in digital audio systems), then it means that at every 20.8 microseconds a new data is obtained in the input, so the filter function has to process the output within this time. Formally, we generate a literal $l_{timing}$ to represent the validity of the time response with the following constraint:

$$l_{timing} \Leftrightarrow ((N \times T) \leq D) \quad (7)$$

where $N$ is the number of cycles spend by the filter, $T$ is the cycle time and $D$ is the deadline.

## IV. EXPERIMENTAL EVALUATION

This section is split into two parts. The experimental setup is described in Section IV-A while Section IV-B describes the results of verifying the digital filters benchmarks using the proposed approach. Note that we do not compare the proposed approach against that presented by [7] since here we model check the actual C code of the digital filters that are intended to be embedded into micro-controllers and DSPs; this is much closer to the real implementation where specific C constructs (e.g., pointer arithmetic and comparisons) are used to implement Eq. (1); and these make the VCs harder.

### A. Experimental Setup

In Table III, we describe some filters chosen with different design types, number of feedback coefficients $N$, number of forward coefficients $M$, input range, and word-length. Note that the column *Bits* indicates the word-length for the integer and fractional parts of the fixed-point numbers including the bit for sign. Note further that the word-length for the fixed-point representation is estimated based on the $\sum|h_k|$ summation and input range in order to obtain optimized filters in terms of reduced number of bits.

For the evaluation of time constraints, we considered the restrictions of a 16 MHz processor operating on a system in which the sample rate is 48 KHz. Note that the sample rate of the system does not interfere in overflow and limit cycle conditions, since this is just a consequence of the fixed-point arithmetic. Here, we used ESBMC v1.21 (which is available at *www.esbmc.org* together with the benchmarks so that other researchers can reproduce the results) and configured it to use the SMT solver Z3 v3.2 [18] with the bit-vector arithmetic enabled since it produces fewer false alarms than integer and real arithmetic (as also observed by Cox et al. [7]). For each benchmark, we invoked the verification engine as follows:

```
esbmc <file> --no-bounds-check --no-pointer-check --no-div-by-zero-check
```

Note that we disable the array bounds, pointer safety, and division by zero assertions since we are interested in checking only filter related properties as previously described in Section III. The above ESBMC call is thus used to check safety properties related to arithmetic underflow and overflow. To check for limit cycle and time constraints, we simply add the options `--function limitCycle` and `--function timing` to the above ESBMC call, respectively.

All the experiments were conducted on an otherwise idle Intel Core i7-2600, 3.40 GHz with 24 GB of RAM running Fedora 64-bits. For all digital filters, the individual time limit has been set to 3600 seconds; the times given were measured using the *time* command.

### B. Experimental Results

After selecting the digital filters, we used their parameters as input to the model implemented in C. Table III summarizes the results obtained for the filters that we verified using ESBMC. It shows the type of failures that we detected on each filter; and we classify them as *OF* for *Overflow*, *LC* for *Limit Cycle*, and *TC* for *Time Constraint* violation. The column *Xsize* shows the number of entries that are applied to the filter, which thus represents the unwinding bound of the program (i.e., the filter function). The verification time (given in seconds) is also shown for each type of failure assertion. Here, TO represents the time-outs (i.e., the tool is aborted after 3600 seconds).

TABLE III. SUMMARY OF RESULTS FOR THE TESTED DIGITAL FILTERS*

| # | Filter** | N | M | $\sum|h_k|$ | Input | Bits | Xsize | Failures | Verification Time (s) | | |
|---|---|---|---|---|---|---|---|---|---|---|---|
| | | | | | | | | | OF | LC | TC |
| 1 | LP-IIR | 2 | 1 | 2 | [-1,1] | <2,4> | 6 | OF, LC | 39 | 4 | <1 |
| 2 | LP-Butterworth-IIR | 3 | 3 | 1.2 | [-1.6,1.6] | <2,5> | 6 | OF | 579 | 634 | <1 |
| 3 | LP-IIR | 3 | 1 | 4 | [-1,1] | <3,4> | 6 | OF, LC | 210 | 29 | <1 |
| 4 | LP-IIR | 3 | 1 | 1.56 | [-1,1] | <2,4> | 6 | - | 110 | 51 | <1 |
| 5 | LP-FIR | 1 | 31 | 1.93 | [-1,1] | <2,6> | 31 | TC | TO | 98 | 1 |
| 6 | HP-ChebyshevI-IIR | 3 | 3 | 1.33 | [-1,1] | <2,10> | 6 | - | 853 | 2058 | <1 |
| 7 | BP-Elliptic-IIR | 3 | 3 | 1.24 | [-1.0,1.0] | <2,10> | 6 | LC | 546 | 474 | <1 |
| 8 | BS-Butterworth-IIR | 3 | 3 | 1.81 | [-1.1,1.1] | <2,8> | 6 | OF | 106 | 1299 | <1 |
| 9 | BP-Elliptic-IIR | 5 | 5 | 0.91 | [-1.1,1.1] | <1,8> | 10 | OF, LC | 7 | 20 | <1 |
| 10 | HP-Butterworth-IIR | 5 | 5 | 1.58 | [-1.27, 1.27] | <2,6> | 10 | LC | 2468 | 1508 | <1 |
| 11 | BP-ChebyshevI-IIR | 5 | 5 | 1.51 | [-1.33, 1.33] | <2,6> | 10 | - | TO | TO | <1 |
| 12 | HP-Elliptic-IIR | 7 | 7 | 5.39 | [-1,1] | <3,13> | 14 | TC | 73 | TO | <1 |

*. Analyzed filters and software are available at www.esbmc.org
**. LP – Lowpass, HP – Highpass, BP – Bandpass, BS – Bandstop

As we can see in Table III, the proposed method can detect failures in digital filters independently of their type, order, or bit-width. However, the verification time tends to be higher for high order filters and for longest word-length formats since these lead to a harder VC, except for the benchmark `HP-Elliptic-IIR` where we can conclude in few seconds that it does not contain any arithmetic underflow and overflow. Note that we time out to check for overflows in digital filter 5 and 11, which contain a high number of forward and feedback coefficients, respectively. Note further that we time out to check for limit cycles in the digital filter 12, which contains the longest word-length of the fractional part. Apart from that, the time constraints are easily verified since it only consists of checking the time response of a sequential piece of code.

V. CONCLUSIONS

In this work, we proposed a new approach to detect failures in fixed-point digital filters using an off-the-shelf bounded model checker. It allows the designer to formally check the given implementation for a specific bit-width and it helps define the word-length to properly represent numbers. In particular, the proposed approach supports the designer to detect problems caused by the finite word-length such as overflows and limit cycles in IIR filters. The experimental results show that we can easily detect failures in low and medium orders digital filters with arbitrary bit-width. However, the verification of high order filters with longest word-length tends to be a hard problem due to the large state space exploration. Additionally, we contributed with a new method based on WCET analysis together with BMC to verify time constraints in digital filters. Since we have modeled and implemented the digital filters in the C language, the proposed approach could also be applied to other existing BMC tools by taking advantage of their robustness and efficiency.